\documentclass[prl,english,showpacs,twocolumn]{revtex4}
\usepackage{graphicx}

\begin{document}

\title{Toward an effective centrality trigger in pp collisions at LHC}

\begin{abstract}
We investigate the impact of very strong small $x$ gluon fields in
colliding nucleons at LHC energies on the interaction of valence
quarks. We find that in the range of small impact parameters, which
contribute significantly to the production of heavy new particles,
several of the valence quarks receive large transverse momenta,
exceeding $1$~GeV/c. This results in a suppression of leading baryon
production and consequently in an additional energy flow to smaller
rapidities. We suggest several triggers for centrality in $pp$
collisions which allow one to study the propagation of partons through
gluon fields of a strength comparable to the ones encountered in heavy
ion collisions at the LHC.
\end{abstract}
\pacs{12.20.-m,  13.85.Hd}
\author{H.J.~Drescher} \affiliation{Frankfurt Institute for Advanced
  Studies, Johann Wolfgang Goethe-Universitat\\ Routh-Moufang-Str. 1,
  60438 Frankfurt am Main, Germany} \author{M.~Strikman}
\affiliation{Department of Physics, Pennsylvania State University
  \\ University Park, PA 16802, USA} 
\date{Dec. 2007} 
\maketitle

It is now widely realized that a fast increase of gluon densities at
small x should result in the onset of the black disk regime (BDR) of
interactions of partons up to rather large virtualities (transverse
momenta). In the BDR a parton gets a large and with energy increasing
transverse momentum when passing through the gluon field ($\sim Q_s$ in
the color glass condensate models, 
for a recent review see \cite{Gelis:2007kn}). In the BDR regime, the parton also loses a
finite fraction of its energy, which allows one to explain the pattern
of leading pion production observed at RHIC in deuteron-gold
collisions\cite{Frankfurt:2007rn}. Gluon densities in the nucleon,
at small transverse distances $\rho$ from the center, are comparable
to those in nuclei at small $\rho$ for the same energies. In
particular for $\rho=0$, and $A\sim 200$, 
\begin{equation} 
{G_N(x, Q^2,\rho=0)\over
  G_A(x, Q^2,\rho=0)} = {2B\over \gamma_A\rho_0 2R_A} \ge 0.5, 
  \label{black}
\end{equation}
where $\rho_0 $ is the nuclear matter density.
Here we took an exponential shape for the gluon distribution in the
nucleon corresponding to the two gluon form factor of $\exp(-Bt/2)$ (a
dipole fit to the two gluon form factor gives an even larger ratio)
and neglected the leading twist gluon shadowing which suppresses
$G_A(\rho)$ by a factor of $\gamma_A(\rho)$.  Since $s_{NN}$ for $pp$
and for heavy ion collisions at the LHC differs by a factor of 6 and
the gluon density depends on $s$ as $s^n, n\sim 0.2$ for the
virtualities of a few GeV$^2$, we find that the gluon densities in
the two cases are similar. 
Hence 
up
 to $\rho \sim 0.6$~fm the gluon
density in a nucleon is comparable to the {\it average} gluon nuclear
density in heavy ion collisions (and 
 larger than the gluon
densities in heavy ion collisions at RHIC).  At larger $\rho$ the
nucleon gluon density decreases rapidly.

There are two challenges: (i) to device a trigger which would select
$pp$ collisions at small impact parameters 
($\left<b\right>\le \left<\rho\right> 
\sim 0.6$~fm) in order to investigate high gluon density effects in $pp$
collisions at densities comparable to $AA$ collisions, and (ii) to
account for high gluon density effects in central $pp$ collisions
which give the dominant contribution to the production of heavy
particles like the Higgs, or in SUSY.

Since the smallest $x_2$ gluons are resolved by large $x$ partons in
the projectile regions - $x_2 \sim 4k_t^2/x s \sim 10^{-7}$ for $x\sim
0.3$ and $k_t=1 \mbox{GeV/c}$ - we expect the strongest effects of the
high density gluon fields for the interaction of valence quarks which
should influence the structure of the final states in the
fragmentation region. Even larger effects are present for the
interaction of the leading gluons due to a factor of 9/4 stronger
interaction in the gluon-gluon channel. These effects are likely to be
manifested at smaller rapidities since gluon distributions have a much
softer $x$ distribution than quarks. (The current MC models usually
neglect hard QCD effects in this kinematic region focusing on hard QCD
dynamics in the central region.)

\begin{figure}[tb]
\begin{center}
\includegraphics[width=0.9\columnwidth]{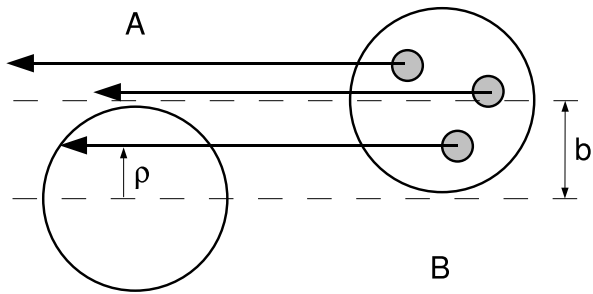}
\caption{ Schematic view of the collision geometry. }
\label{fig:mult}
\end{center}
\end{figure}

For a $pp$ collision at a given impact parameter $b$, individual
partons of one nucleon can pass at a range of transverse
distances $\rho$ from the second nucleon and hence encounter
significantly different local gluon densities (see
Fig.\ref{fig:mult}). In this paper we analyze the effects of the
valence quark interaction with small x gluon fields taking into
account the geometry of the collisions. This will allow us to
determine how frequently valence quarks in $pp$ collisions at
different impact parameters $b$, experience hard collisions in which
they obtain a large transverse momentum. Based on this study we
propose a series of centrality triggers which allow to select
collisions at much smaller impact parameters than in generic inelastic
events and hence will provide an opportunity to study the high gluon
field effects in $pp$ collisions.  We also suggest that the $pp$
collisions leading to production of new particles like the Higgs boson
should be accompanied by a 
significantly
stronger flow of energy from the
fragmentation regions to smaller rapidities than in generic inelastic
collisions.

{\it Description of the model.}  To model the fragmentation region in
$pp$ collisions we take a simple model for the three quark wave
function with the distribution of quarks over transverse distance from
the center given by
$\exp(-A\rho_i^2)$ with $<\rho^2> \sim 0.3 fm^2$ matched to describe
the distribution of the valence quarks as given by the axial nucleon
form factor. Accordingly, the event generator produces the values of
$\rho_i$ for three quarks which are not correlated. Note that one does
not expect a very strong correlation between $\rho$'s due to the
presence of additional partons in the wave function (gluons, $q\bar q$
pairs). Nevertheless we checked that a requirement $| \sum_{i=1}^3 {\vec
  \rho}_i | \le 0.1$~fm does not change results noticeably. Hence we
neglect possible correlations in $\rho$ between valence quarks.  We
also assume that there are no significant transverse correlations
between small x ($x\sim 10^{-5}$) partons. This assumption is based on
the presence of diffusion in $\rho$ in the small x evolution which
should wash away whatever correlations may be present at large x ($x>
0.1$). (For a discussion of the evidence for such correlations see
\cite{annual}.)

When computing the momentum fractions of the quarks, we need to know
the virtuality at which the quarks are resolved. Since the latter
quantity is not known beforehand, we generate $x_{B,i}$ and
$\vec{\rho_i}$ from $dx/x=const.$ and $d\rho=const.$
distributions. The selection according to the structure functions and
the form factor is done in the end, after specifying $Q_s^2$, via
rejection.  For a given impact parameter $\vec{b}$ and relative
positions of the quarks $\vec{\rho_i}$ in the projectile, we estimate
the density within the color glass condensate approach. So we need to
know the saturation scale in the target for the three valence quarks.

\begin{figure}[tb]
\begin{center}
\includegraphics[width=0.9\columnwidth]{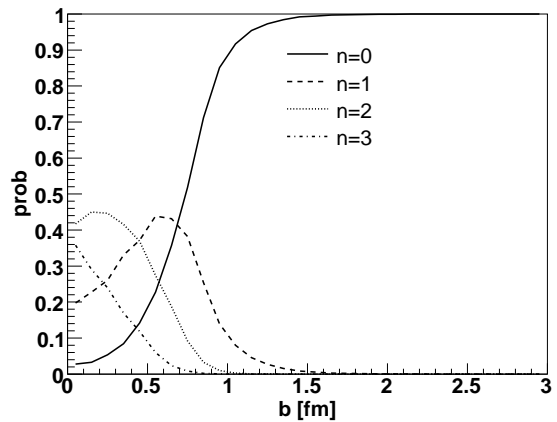}
\caption{Probability for the different classes of
  events with $n$ quarks struck at a given impact parameter $b$.}
\label{fig:mult1}
\end{center}
\end{figure}

\begin{equation}
Q^{2}=Q_{s}^{2}(x_{A},\vec{|b}+\vec{\rho_i}|),
\end{equation}
with $x_A= Q^2/(s x_B)$. The saturation momentum is parameterized as 
\begin{equation}
Q_{s}^{2}(x_{A},\rho)=Q_{s,0}^{2}\left(\frac{x_{0}}{x_{A}}\right)^{\lambda}F_{g}(x_{A},\rho;Q_s^{2})/c_{F},\label{eq:Qsat2}
\end{equation}
where $c_{F}$ normalizes the density. We choose $x_0=0.01$,
$Q_{s,0}^2=0.6$~GeV$^2$ and $c_{F}=F_{g}(x_{0},0;Q_{s,0}^{2})$ such
that the saturation momentum in the center of the target at $x_A=x_0$
is just $Q_{s,0}^{2}$. The implicit definition for the saturation
scale in eq. (\ref{eq:Qsat2}) is solved by a simple iteration, the
expression converges after a few steps.  Finally, the whole
configuration is accepted with the probability
\begin{equation}
p \sim \rho F_{g}(x_{B},\rho;Q_{s}^{2})x_{B}f_{\mbox{GRV}}(x_{B},Q_{s}^{2})~,
\end{equation}
where $x f_{\rm GRV}$ are standard GRV structure functions of the
proton, and the two-gluon form factor at high momentum fraction $x_B$
describes the spatial distribution of the valence quarks.  The actual
transverse momentum kick is then drawn from the distribution
\cite{Boer:2002ij,Drescher:2005ak}
\begin{equation}
C(k_{t})\sim\frac{1}{Q_{s}^{2}\log\frac{Q_{s}}{\Lambda_{QCD}}}\exp(-\frac{\pi
  k_{t}^{2}}{Q_{s}^{2}\log\frac{Q_{s}}{\Lambda_{QCD}}})\,.
\end{equation}
We conservatively considered only the case when the BDR is reached for
$Q_s\ge 1$~GeV/c and counted only quark interactions in which the
quark received a transverse momentum $k_t\ge 0.75 \mbox{GeV/c}$.  The
reason for such a cut is that for such momenta, the probability to
form a nucleon with large longitudinal momentum is suppressed, as a
minimum, by the square of the nucleon form factor $F_N^2(k_t)$. In
the BDR a quark not only gets a large transverse momentum but also
loses a finite fraction of its momentum - we neglected this effect in
our analysis since it is likely to be on the scale of a 10\% energy
loss and would not affect our qualitative discussion.  We also assumed
in the analysis that the gluon densities at small x are not
fluctuating significantly for a impact parameters where the BDR holds
(this is an assumption usually made in the saturation models where the
the small gluon field is treated as nearly classical field).
Neglecting fluctuations seems safe for the small $x$ in our analysis,
though for larger $x$ relevant at the Tevatron collider 
they
 may be more
important.

\begin{figure}[tb]
\begin{center}
\includegraphics[width=0.9\columnwidth]{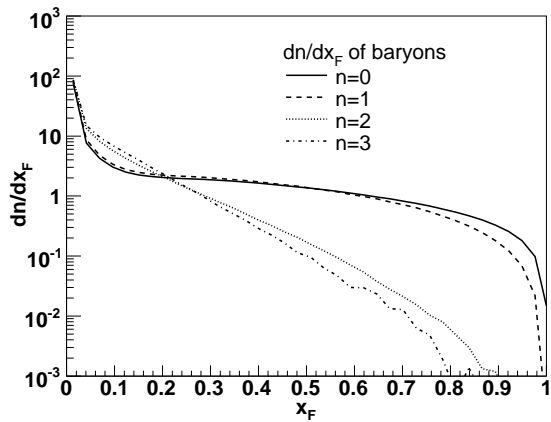}
\caption{ $dn/dx_F$ distribution of baryons. }
\label{fig:xF}
\end{center}
\end{figure}

As a first result, the probabilities $P_i$ for $i$ quarks to interact
in the BDR regime (number of "wounded" quarks, $N_w$) are presented in
Fig.~\ref{fig:mult1} as a function of the impact parameter $b$. Note
that for small $b$ interactions $N_w\ge 2$ dominate. In such
interactions each quark fragments independently, so the spectra for
$N_w=2$ and $N_w=3$ should be rather similar and shifted to much
smaller $x_F$ than in soft interactions where the spectra of nucleons
are known to be flat in $x_F$ in a wide range of $x_F$.  At the same
time for $N_w=1$, the dominant process is likely to be diquark
fragmentation which is similar to the soft mechanism.  Hence one
expects two distinctive shapes of $x_F$ distributions - one with
leading baryons ($N_w=0,1$) and another with suppressed forward
scattering ($N_w=2,3$). We implemented the fragmentation of the system
produced in the first stage by constructing strings which decay using
the LUND method. There are always two strings, drawn between a quark
and a diquark from the interacting particles. When a quark of the
diquark receives a high transverse momentum, the diquark becomes a
system of two quarks and a junction.  This has the nice property that
one recovers the diquark when the invariant mass between the two
quarks is small.  The results are plotted in Fig. \ref{fig:xF} and are
in good agreement with the qualitative expectations discussed above.

The total probability of an inelastic interaction at given impact
parameter can be expressed through the impact factor $\Gamma(b)$,
which is the Fourier transform of the elastic amplitude, as
$1-\left|1-\Gamma(b)\right|^2$. On the other hand in our model we
calculate the probability of inelastic interactions due to the hard
interaction of $i$ quarks of one of the nucleons, $P_i$.  Since in the
discussed approximation there is no correlation between the
interactions in two fragmentation regions for a fixed $b$ the
probability that none of six valence quarks will get a kick is $1-
(1-P_1-P_2-P_3)^2$. For a consistency of the model we need that
\begin{equation} 
1-\left|1-\Gamma(b)\right|^2 \ge 1-  (1-P_1-P_2-P_3)^2,
\end{equation}
or 
\begin{equation} 
\Gamma(b) \ge P_1+P_2+P_3.
\end{equation}

\begin{figure}[tb]
\begin{center}
\includegraphics[width=0.9\columnwidth]{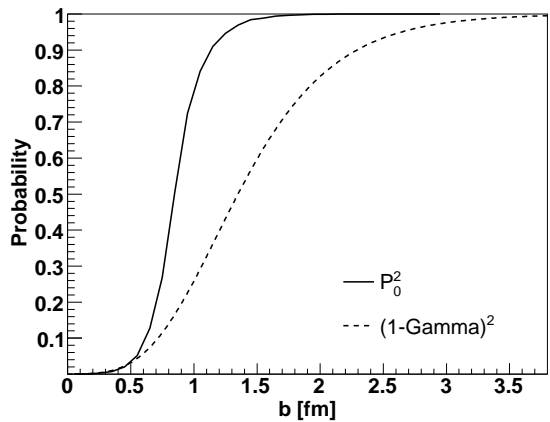}
\caption{Probability that no quarks were struck (solid curve as compared to the probability of absence of generic inelastic interaction (dashed curve) as a function of $b$.}
\label{fig:mult2}
\end{center}
\end{figure}

We find that this condition is 
satisfied for all $b$ (within the uncertainties of the current models for pp elastic scattering at LHC) - see Fig.~\ref{fig:mult2} indicating that hard collisions lead to 
 saturation of  the black limit for $\Gamma$ for small
$b$ at LHC energies, while at large $b$ hard interactions give a very
small contribution. Hence the model provides a dynamic explanation of $\Gamma (b\sim 0)=1$ and leads  to expansion of the $\Gamma(b)=1$ region with increase of energy.


Very strong dependence of $N_w$ on $b$ and a strong correlation of
$N_w$ with the multiplicity of leading baryons allows one to determine
the effectiveness of a centrality trigger based on a veto for the
production of leading baryons with $x>x_{tr}$ as a function of
$x_{tr}$. We find than an optimal value of $x_{tr}$ is $\sim
0.1$. Current configurations of several LHC detectors allow to veto
neutron production in this x-range. TOTEM, in addition, allows to veto
production of protons with $x_F>0.8$.  Since neutron and proton
multiplicities are similar, a one side veto for production of both
charged and neutral baryons leads approximately to the same result as
a two side veto for neutron production.  Accordingly we will give
results both for single side veto and for two side veto for both
neutral and charged baryons (understanding that the full
implementation of the latter option would require certain upgrades of
the detectors some of which are currently under discussion).

\begin{figure}[t]
\begin{center}
\includegraphics[width=0.9\columnwidth]{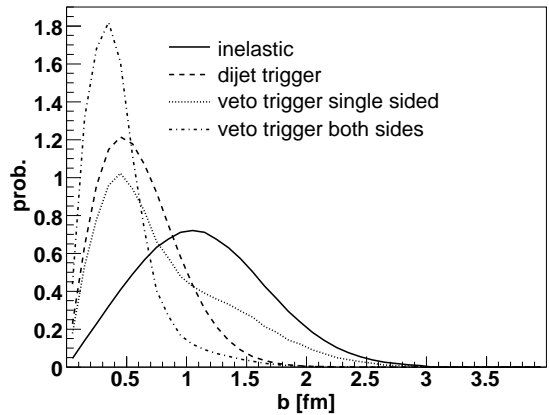}
\caption{Impact parameter distributions for inelastic events, the dijet trigger and single and double sided veto-trigger (no baryon in the region $x_F> 0.1$).}
\label{fig:trigger}
\end{center}
\end{figure}

The results of the calculations are presented in
Fig.~\ref{fig:trigger} together with the distribution over $b$ for
generic inelastic events and the central dijet trigger
\cite{Frankfurt:2003td}. We see that the single side veto trigger
leads to a centrality similar to that of a the dijet trigger, while a
double side veto leads to the most narrow distribution in $b$.  An easy
way to check this expectation would be to compare other
characteristics of events due to these triggers - one expects for
example a progressive increase of the central multiplicity with
a decrease of average $b$.
 
\begin{figure}[t]
\begin{center}
\includegraphics[width=0.9\columnwidth]{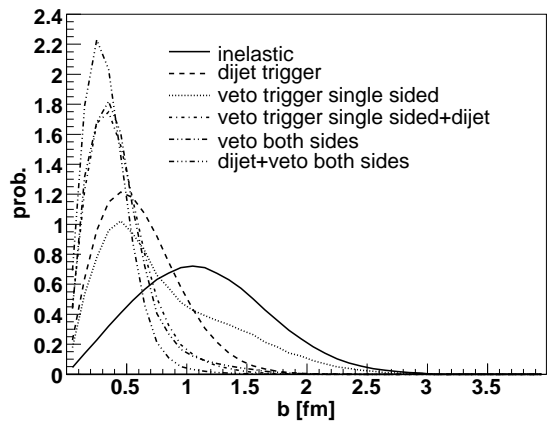}
\caption{The combination of dijet and veto trigger gives the best
  constraints on central events in $pp$-collisions.}
\label{fig:combined}
\end{center}
\end{figure}
The most narrow distributions can be achieved by combining two types
of the triggers - veto and dijet - Fig.~\ref{fig:combined}.  We find
that in this case we reach the limit that 
$\left<\rho_{tr}\right> =
\left(\left< \rho^2\right>
 +\left< b^2\right>\right)^{1/2} $  
 becomes
comparable to $\left<\rho\right>$ which is the smallest possible average $\left<\rho\right>$ for $pp$ or DIS collisions.  

In line with discussion of  Eq.~\ref{black} we expect  that for our best centrality triggers average gluon densities will be comparable to those in the heavy ion collisions.
Indeed, we estimate the average gluon density encountered by the leading
partons, and find it larger than the one encountered at central impact
parameters in pA collisions at RHIC where BDR holds at least for
$p_t\le 1.5 \mbox{GeV/c}$. Further, it is about $\frac{1}{2}$ of the
average gluon density in heavy ion collisions at LHC. (However
dispersion in the value of $Q_s$ is much larger in the $pp$ case).

We also estimated the effect of a dijet trigger (which is the same
selection of impact parameters as for production of heavy new
particles). We find that in 16\% of these events at least in one
fragmentation region no leading baryons would be present with $x_F\ge
0.1$ as compared to a generic case where this fraction is 8\%.  This
should result in a significantly larger flow of energy to central
rapidities and strong fluctuations of this flow on an event by event
basis.  Since the probability of interaction for gluons with $x\ge
10^{-2}$ is at least as large as for quarks with $x>0.1$ we expect
that many of the gluons with these $x$ will also lose coherence with
the low transverse momentum partons and will fragment independently
further contributing to the energy flow to central rapidities. It is
important to investigate how these effect would impact on the various
triggers suggested for the searches of the new particles.

In conclusion, we have demonstrated that it will be feasible to
trigger on very central $pp$ collisions at the LHC. The high gluon
densities encountered by the colliding partons in such collisions
facilitate studies of novel QCD effects complementary to heavy ion
collisions, such as the $p_t$ broadening of the $Z$-boson distribution
as compared to generic events, multiplicity fluctuations, elliptic
flow, etc.

We thank L.Frankfurt and C.Weiss for numerous discussions. We also
benefited from discussions of the experimental capabilities and plans
of upgrades of ATLAS, CMS and TOTEM detectors with V.Avati, J.Bjorken,
K.Eggert, H.Jung, A. De Roeck, and S.White.
This work is supported by BMBF grant 05~CU5RI1/3

\end{document}